\documentclass{nature1}
\usepackage{graphicx}
\usepackage{hyperref}

\title{Observations of an Extreme Storm in Interplanetary Space Caused by Successive Coronal Mass Ejections}

\author{Ying D. Liu$^{1}$, Janet G. Luhmann$^{2}$, Primo\v{z} Kajdi\v{c}$^{3,4}$, Emilia K. J. Kilpua$^{5}$, No\'{e} Lugaz$^{6}$, Nariaki V. Nitta$^{7}$, Christian M\"{o}stl$^{8,9}$, Benoit Lavraud$^{3,4}$, Stuart D. Bale$^{2}$, Charles J. Farrugia$^{6}$ \& Antoinette B. Galvin$^{6}$}

\begin{document}

\maketitle

\begin{affiliations}

 \item State Key Laboratory of Space Weather, National Space Science Center, Chinese Academy of Sciences, Beijing, China.
 
 \item Space Sciences Laboratory, University of California, Berkeley, CA 94720, USA.
 
 \item Universit\'{e} de Toulouse, UPS-OMP, IRAP, Toulouse, France.
 
 \item CNRS, IRAP, 9 Av. colonel Roche, BP 44346, F-31028 Toulouse cedex 4, France.
  
 \item Department of Physics, University of Helsinki, Helsinki, Finland.
 
 \item Space Science Center, University of New Hampshire, Durham, NH 03824, USA.
 
 \item Lockheed-Martin Solar and Astrophysics Laboratory, Palo Alto, CA 94304, USA.
 
 \item Institute of Physics, University of Graz, Graz, Austria.
 
 \item Space Research Institute, Austrian Academy of Sciences, 8042 Graz, Austria.
 
Correspondence and requests for materials should be addressed to Y. Liu (email: liuxying@spaceweather.ac.cn).

\end{affiliations}

\begin{abstract}

Space weather refers to dynamic conditions on the Sun and in the space environment of the Earth, which are often driven by solar eruptions and their subsequent interplanetary disturbances. It has been unclear how an extreme space weather storm forms and how severe it can be. Here we report and investigate an extreme event with multi-point remote-sensing and in-situ observations. The formation of the extreme storm showed striking novel features. We suggest that the in-transit interaction between two closely launched coronal mass ejections resulted in the extreme enhancement of the ejecta magnetic field observed near 1 AU at STEREO A. The fast transit to STEREO A (in only 18.6 hours), or the unusually weak deceleration of the event, was caused by the preconditioning of the upstream solar wind by an earlier solar eruption. These results provide a new view crucial to solar physics and space weather as to how an extreme space weather event can arise from a combination of solar eruptions.

\end{abstract}

While coronal mass ejections (CMEs), large-scale expulsions of plasma and magnetic field from the solar corona, have been recognised as drivers of major space weather effects, some fundamental questions for solar physics and space weather have not been addressed. Specifically, how does an extreme space weather storm form and evolve, and how severe can it be especially when it reaches the Earth? 

An extreme space weather storm is termed a low-probability, high-consequence event, otherwise called a solar superstorm\cite{report08,cannon13}. The largest known example of extreme space weather before the space era is the 1859 Carrington event\cite{carrington1859}. It produced a geomagnetic storm, whose minimum $D_{\rm st}$ is estimated\cite{siscoe06} to be approximately $-850$ nT. The $D_{\rm st}$ index is a measure of the severity of geomagnetic storms (the more negative the more intense). The most severe geomagnetic storm of the space age has a minimum $D_{\rm st}$ of $-548$ nT and was caused by the 1989 March 13 solar wind disturbance\cite{cliver04}, which led to the collapse of Canada's Hydro-Qu\'{e}bec power grid and a resulting loss of electricity to six million people for up to nine hours. Therefore, understanding extreme space weather is also vital to the modern society and its technological infrastructures. 

A statistical analysis\cite{yashiro04} suggests that CME speeds near the Sun range from several tens to about 3000 km s$^{-1}$. In particular, the limit of 3000 km s$^{-1}$ is argued to indicate an upper bound to the maximum energy available to a CME from solar active regions\cite{gopal05}. Most CMEs, however, end up with speeds and magnetic fields comparable to those of the ambient solar wind when they reach the Earth, i.e., about 450 km s$^{-1}$ and 10 nT, respectively\cite{richardson10,liu05}. This implies that, while the active region may decide how energetic a CME can be near the Sun\cite{gopal05}, the subsequent evolution in interplanetary space is also crucial in determining its severity at the Earth. How decisive the interplanetary evolution is remains unclear. 

A new view of how extreme space weather can be created by CME-CME interactions in interplanetary space is discovered from the 2012 July 23 event, which had a maximum speed of $3050\pm260$ km s$^{-1}$ near the Sun. This event, with complete modern remote-sensing and in-situ observations from multiple vantage points, provides an unprecedented opportunity to study the cause (and perhaps consequences as well) of extreme space weather. We illustrate how successive CMEs can be enhanced into a superstorm as they interact en route from the Sun to 1 AU. Note that, while there are many parameters that could be used to measure the extreme or severity of a solar storm\cite{cliver04, riley12, schrijver12}, here we focus on the solar wind transient speed ($v$) and magnetic field ($B$) at a distance of $\sim$1 AU. This is because their cross product, the dawn-dusk electric field (see Methods), controls the rate of the solar wind energy coupling to the terrestrial magnetosphere\cite{dungey61}. The CME-CME interactions studied here produced an extremely high solar wind speed of $2246\pm110$ km s$^{-1}$ and an unusually strong ejecta magnetic field of $109\pm1$ nT observed near 1 AU at STEREO A. This record solar wind speed and magnetic field would have generated the most severe geomagnetic storm since the beginning of the space era, if the event had hit the Earth. The extreme space weather generation scenario is similar to the case of a ``perfect storm", a phrase which was initially coined for the 1991 October Atlantic storm but now describes any event where a rare combination of circumstances will aggravate a situation drastically. 

\noindent\textbf{Results}

\noindent\textbf{Multi-point imaging observations.} Figure~1 shows the positions of three widely separated spacecraft in the ecliptic plane, including the Solar and Heliospheric Observatory (SOHO)\cite{domingo95} upstream of the Earth and the Solar Terrestrial Relations Observatory (STEREO)\cite{kaiser08}. STEREO comprises two spacecraft with one preceding the Earth (STEREO A) and the other trailing behind (STEREO B). During the time of the solar eruptions on 2012 July 23, STEREO A and B were 121.3$^{\circ}$ west and 114.8$^{\circ}$ east of the Earth with a distance of 0.96 AU and 1.02 AU from the Sun, respectively. The event originated from sunspot group NOAA 11520 (S15$^{\circ}$W133$^{\circ}$), so the source location on the Sun was 43$^{\circ}$ behind the west limb from the point of view of the Earth, whereas about 12$^{\circ}$ west of the central meridian as viewed from STEREO A and about 22$^{\circ}$ behind the east limb for STEREO B. Obviously STEREO A gives the best observations of the activity on the solar disk, while STEREO B and Earth (SOHO) provide the best views of the associated CMEs. In particular, STEREO B provides a view of the low coronal activity with minimised projection effects.

The event started with two consecutive prominence eruptions separated by about 10 - 15 minutes, as shown in Figure~2a-c (also see Supplementary Movie 1). At 02:27 UT (Fig.~2b), the first erupting prominence (E1) was not visible any more, but a CME (CME1) appeared in COR1 of STEREO B. The second erupting prominence (E2) was still rising and produced a bulging of the coronal structure just south of CME1, which was later recognised as the south edge of the second CME (CME2). The north edge of CME2 was not clear and likely overlapped with CME1. The erupting prominence E2 (Fig.~2c), however, was even visible in the field of view (FOV) of COR1 and overlapped with a structure with enhanced brightness, so the erupting prominence E2 at 02:37 UT probably marked the core of CME2. At about the same time (02:36 UT), the leading edge of CME1 had already propagated beyond the FOV of COR1 (Fig.~2c). The second eruption appeared stronger than the first one. In COR2 of STEREO B (Fig.~2d, e, Supplementary Movie 2), definite twin-CME signatures were lost presumably due to the CME-CME interaction, although a concave-outward structure may give an impression of two CMEs interweaved together. A shock signature, indicated by a weak edge ahead of the CME front\cite{vourlidas03}, was already developed. COR2 aboard STEREO A (Fig.~2g, h, Supplementary Movie 3) first saw material largely west of the Sun at 02:54 UT, but later the event was almost east-west symmetric, which may indicate a change in the longitudinal propagation direction. SOHO gave another side view of the event (Fig.~2f, i), but did not see clear twin-CME signatures. This is probably due to the low time resolution of LASCO (12 minutes) and a missing image at 03:00 UT for LASCO/C2. The two CMEs occurred so close that one could also argue for a complex single event composed of two eruptions.  

Figure~3 shows the time-elongation maps, which are produced by stacking the running difference images within a slit along the ecliptic plane (here we take a position angle of $90^{\circ}$ measured counterclockwise from the solar north for STEREO B while clockwise for SOHO)\cite{sheeley08, davies09, liu10a}. The time-elongation map from COR1 of STEREO B shows two adjacent tracks (Fig.~3b), which agrees with the observations of two closely launched CMEs (Fig.~2a-c, Supplementary Movie 1). These two tracks seem to merge in the map from COR2 of STEREO B. The time-elongation map from LASCO/C2 also appears to have two adjacent tracks around 03:00 UT (Fig.~3a), but due to the low time resolution and a missing image at 03:00 UT the correspondence between SOHO and STEREO B cannot be established for the second track. 

\noindent\textbf{CME kinematics and flare EUV flux.} We determine the kinematics of the CME leading front with a triangulation technique\cite{liu10a,liu10b}. This technique has no free parameters and has had success in tracking interplanetary propagation of various CMEs\cite{liu11, liu12, liu13, mostl10, harrison12, temmer12, mishra13}. STEREO B and SOHO provide the best elongation measurements owing to their vantage points, as shown in Figure~3. Elongation angles are extracted from the leading track in the time-elongation maps, as there is no ambiguity in identifying the correspondence between SOHO and STEREO B for the leading track. The elongation angles are then used as input to the triangulation technique. Uncertainties in the elongation measurements can be used to evaluate the errors in the propagation direction, radial distance and speed of the CME front\cite{liu10a,liu10b}. 

The resulting kinematics in the ecliptic plane are displayed in Figure~4. The propagation direction is converted to an angle with respect to the Sun-STEREO A line. If the angle is positive (negative), the CME front would be propagating west (east) of the Sun-STEREO A line. The propagation angle started from about the solar source longitude and then showed a transition toward the Sun-STEREO A line. The change in the propagation direction occurred at 02:42 - 03:05 UT when the CME front reached about 6 - 11 solar radii. This is consistent with the CME images seen by COR2 of STEREO A (Fig.~2g, h). The change in the propagation angle had a good timing with, and was thus probably caused by, the CME-CME interaction during their merging. The amount of the longitudinal deflection is about 10$^{\circ}$, similar to a previous CME-CME interaction case\cite{lugaz12}. The speed of the CME leading edge first increased to about 3050 km s$^{-1}$ around 03:14 UT at a distance of about 13 solar radii, and then decreased to about 2700 km s$^{-1}$.

Measurements of X-ray emissions from the associated flare are not available, so we use the full-disk EUV flux at 195 \AA\ observed by STEREO A as a proxy (Fig.~4c). The EUV flux at 195 \AA\ is proved to be a good tracer of the X-ray emission in the 1 - 8 \AA\ band; the magnitude of the flare associated with the 2012 July 23 event is at most X2.5 estimated from the relationship between the EUV and X-ray fluxes\cite{nitta13}. The flare is not particularly intense compared with the Carrington ($>$X10), 2003 October 28 ($>$X17) and 2003 November 4 (X28) flares\cite{cliver04}, which is surprising given how extreme the solar wind disturbance was. This raises an intriguing and important question of how the event became extreme in interplanetary space. The EUV flux shows two peaks with a time separation of about 15 minutes. This is consistent with the two consecutive eruptions seen in the EUV images at 304 \AA\ from STEREO B (Fig.~2a-c). The CME peak speed was about 30 minutes after the maximum EUV flux, which may suggest further acceleration post the flare maximum\cite{liu13}. 

\noindent\textbf{In-situ measurements of the extreme storm.} Figure~5 shows the in-situ measurements from the PLASTIC and IMPACT instruments aboard STEREO A at 0.96 AU. A forward shock passed STEREO A at 20:55 UT on July 23, with a transit time of only 18.6 hours (if we take 02:20 UT as the CME launch time). Two interplanetary CMEs (ICMEs), in-situ counterpart of CMEs, can be identified from the magnetic field measurements behind the shock. Across the shock and ICMEs the proton density and temperature measurements are largely missing here because the conditions exceeded the nominal detector ranges. We use the density of electrons with energies above 45 eV (multiplied by a factor of 5) as a proxy of the plasma density. It matches the proton density well when the proton data are available during the time interval. The solar wind speed jumps from about 900 to 2246 km s$^{-1}$ across the shock, which is a remarkably high increase. Both the post-shock peak speed  (2246 km s$^{-1}$) and the peak magnetic field strength (109 nT) in the early part of the ejecta are among the few largest on record ever measured near 1 AU. Extremely fast solar wind with a speed of $\sim$2000 km s$^{-1}$ was measured near 1 AU on two occasions previously, the 1972 August 4-5 and 2003 October 29-30 events\cite{duston77, skoug04}. The peak magnetic field associated with the two previous events is about 110 and 68 nT, respectively, which occurred only briefly in the sheath region between the shock and ejecta. The present case, however, has the maximum magnetic field taking place in the first ICME (ICME1), and the ejecta magnetic fields larger than 60 nT lasted for over 6 hours. 

The magnetic field components are given in \textbf{RTN} coordinates, where ${\bf R}$ points from the Sun to the spacecraft, ${\bf T}$ is parallel to the solar equatorial plane and along the planet motion direction, and ${\bf N}$ completes the right-handed system. The two ICMEs seem to have quite different magnetic configurations. Reconstruction assuming a flux-rope geometry\cite{hau99,hu02} gives a right-handed structure for both of the ICMEs. Their axis orientations, however, are very oblique with respect to each other: an elevation angle of about $69^{\circ}$ and azimuthal angle of about $226^{\circ}$ (in \textbf{RTN} coordinates) for ICME1, and an elevation angle of about $-44^{\circ}$ and azimuthal angle of about $186^{\circ}$ for ICME2. These axis orientations should be considered as rough estimates, as we do not have good density and temperature measurements and the magnetic field structures of the two ICMEs are irregular. 

The magnetic field has a sustained southward component (negative $B_{\rm N}$), which was larger than 20 nT for more than 5 hours (with a maximum value of 52 nT inside ICME2). The extremely high solar wind speed combined with the prolonged strong southward field component would produce an extremely severe geomagnetic storm if the event hit the Earth. We evalute the $D_{\rm st}$ index using two empirical formulas\cite{burton75, om00}, based on the solar wind measurements (see Methods). The resulting $D_{\rm st}$ profiles show a classic geomagnetic storm sequence: a sudden commencement caused by the shock, a main phase, and then a recovery phase. The difference between the two $D_{\rm st}$ profiles is owing to different assumptions on the decay of the terrestrial ring current in the two formulas (see Methods). ICME1 does not produce negative $D_{\rm st}$ values because of its largely northward magnetic fields. The minimum $D_{\rm st}$ value is $-1150$ and $-600$ nT, respectively. This is more intense than the most severe geomagnetic storm of the space era caused by the 1989 March event\cite{cliver04}. The modeled minimum $D_{\rm st}$ is comparable to that of the 1859 Carrington event\cite{siscoe06}, the largest known solar disturbance before the space age. These results are consistent with previous estimates or speculations\cite{russell13, baker13, ngwira13}. 

The average transit speed is about 2150 km s$^{-1}$, obtained by dividing the distance of STEREO A (0.96 AU) by 18.6 hours. This is comparable to the peak solar wind speed at STEREO A, which implies that the event was not slowed down much by the ambient medium. Comparing the CME speed near the Sun (3050 km s$^{-1}$) with the peak solar wind speed at STEREO A gives a slowdown of only about 800 km s$^{-1}$. Since CME deceleration is due to interactions with the ambient solar wind, an initially faster CME will have a stronger deceleration\cite{gopal00, liu11}. The minimal slowdown of the present case as well as the extremely strong magnetic field is surprising, and raises a question of how the speed and internal magnetic field of the event were maintained or achieved during transit from the Sun to 1 AU.

\noindent\textbf{Generation of the extreme speed and magnetic field.} The cause of the minimal slowdown of the 2012 July 23 event is of particular interest. Figure~6 shows an ICME (ICME0) preceding the July 23-24 event. Two forward shocks are also observed ahead of ICME0, which may indicate a series of eruptions on the Sun before the July 23 event. ICME0 was probably produced by a CME launched around 05:30 UT on July 19 with a peak speed about 1600 km s$^{-1}$ from the same active region (AR 11520) as the July 23 event. The region trailing ICME0 has a density as low as 1 cm$^{-3}$. This low density was likely caused by, first, removal of some of the solar wind plasma by the series of preceding eruptions including ICME0, and second, further rarefaction by the fast motion of ICME0. The magnetic field in the same region predominantly lies along the radial direction from the Sun, presumably due to stretching of the post-event field by ICME0. These observations indicate that the July 23 complex event was moving through a density depletion region with radial magnetic fields. The significantly reduced solar wind drag and field line tension force can explain the marginal slowdown of the July 23 event. 

Combination of the imaging observations with in-situ measurements suggests that the mechanism of creating the extremely strong magnetic field is CME-CME interaction. The time interval of ICME1 is very short (Fig.~5), probably owing to compression by ICME2 from behind. This compression started from the nascent stage of the CMEs (Fig.~2), and continued all the way to large distances as revealed by a speed drop in the interaction interface between the two ICMEs (Fig.~5b). A possible scenario is that a shock driven by CME2 was first overtaking CME1 from behind before the direct compression by CME2 occurred. The shock would strengthen the magnetic field inside CME1 as well when it was propagating through and compressing CME1\cite{liu12,mostl12,lugaz05}. The magnetic field inside CME2 would also appear enhanced because the expansion of CME2 was inhibited by the CME-CME interaction.

Information from solar wind suprathermal electrons supports the picture of CME-CME interaction, as shown in Figure~7. The pitch angle (PA; the angle between the electron motion direction and the magnetic field) distribution of the electrons and their heat flux (thermal energy flux carried by the electrons) parallel to the magnetic field trace the magnetic field topology very well: whenever $B_R$ switches between negative (sunward) and positive (anti-sunward), the PAs flip between 180$^{\circ}$ and 0$^{\circ}$ and the parallel heat flux changes between negative and positive correspondingly. This indicates that the electron beams always move away from the Sun along the magnetic field. Bidirectional streaming electrons (electrons that exhibit peak flux both parallel and anti-parallel to the magnetic field), a typical signature of ICMEs, are not observed during the time period of Figure~7. The PA distribution becomes irregular in the region between the two ICMEs, confirming its identification of the interaction interface. At 01:51 UT on July 24 the magnetic field components and the parallel heat flux showed an abrupt change. This is probably the heliospheric current sheet entrained in the CME-CME interaction region. The largest heat flux is seen in the sheath region between the shock and ejecta, presumably owing to shock compression. The heat flux is also enhanced in ICME1 and the interaction interface but decreases into ICME2, which is consistent with the situation that ICME2 was compressing ICME1 from behind. (There are a few times when the perpendicular heat flux is comparable to the parallel one, which raised a concern that some of the data might be contaminated. We decided to show the heat flux data, as there could also be a localized mechanism associated with the CME-CME interaction that breaks down the gyrotropic motion of the particles with respect to the magnetic field.)

The above observations indicate how an almost ``perfect" combination of coronal and interplanetary conditions could produce an extreme solar wind disturbance near 1 AU. First, an earlier large CME should occur at the right time in order to have a minimal deceleration of the later CMEs. It should occur neither too early (otherwise there would be no solar wind rarefaction or magnetic field line stretching upstream of the later ones), nor too late so that it would not slow down the later ones. Second, the later CMEs should be launched in quick succession. They interact as close to the Sun as possible, so their expansion would be inhibited after the eruptions. The fast transit to 1 AU due to the preconditioning of the upstream solar wind by the earlier CME would significantly reduce the time for them to expand as well. An explicit assumption is that the propagation directions of these CMEs are close to each other. An event with extremely enhanced internal magnetic field and extremely high solar wind speed near 1 AU can be generated under these circumstances besides the initial active region conditions, as we have seen from the 2012 July 23 event. 

\noindent\textbf{Discussion}

The present results show how interactions between consecutive CMEs resulted in a ``perfect storm" near 1 AU, i.e., nonlinear amplification of the events into an extreme one. At least three effects of CME-CME interactions are discovered from the 2012 July 23 event: a change in the propagation direction toward a head-on impact with STEREO A, an extreme enhancement of the ejecta magnetic field, and an exceptionally modest deceleration. Combined together, these effects caused a record solar storm near 1 AU since the dawn of the space era. Had the event hit the Earth, it would have produced a record geomagnetic storm with a minimum $D_{\rm st}$ of $-$1150 - $-$600 nT. This new view on the generation of extreme space weather, especially how a magnetic field larger than 100 nT was produced inside an ICME near 1 AU and preconditioning of the heliosphere for minimal CME deceleration, emphasizes the crucial importance of CME-CME interactions in space weather research and forecasting as previously postulated\cite{gopalswamy01, burlaga02, farrugia04, lugaz05, liu12, liu13}.

These results can serve as a benchmark for space weather prediction models, especially for testing the high-end distributions of the speed and magnetic field of solar wind transients of which detailed observations have been extremely lacking. A preliminary MHD modeling indicates that, while some of the features described here can be captured (i.e., rarefaction and field line stretching by preceding eruptions), the solar wind speed and magnetic field near 1 AU are significantly underestimated. Clearly, to create a realistic simulation that can reproduce the key space weather elements the physical processes associated with the CME-CME interactions have to be taken into account and properly treated. Also note that the 2012 July 23 superstorm occurred in a historically weak solar cycle. Observations of such a solar superstorm during a very weak solar cycle indicate that extreme events are not as infrequent as we imagine. Only with multi-point remote-sensing and in-situ observations can we catch the event in its entirety, obtain a complete view of what was happening around the Sun and in the helliosphere, and discover for the first time how a combination of coronal and interplanetary conditions produced such an extreme event near 1 AU. These would have far-reaching implications for solar and heliospheric physics as well as space weather research and forecasting. 

\noindent\textbf{Methods}

\noindent\textbf{Estimate of the $D_{\rm st}$ index.} We calculate the $D_{\rm st}$ values using the formulas of Burton et al.\cite{burton75} and O'Brien \& McPherron\cite{om00}, respectively. Both of the two algorithms assume the following form
\begin{equation}
D_{\rm st} = D_{\rm st}^{\ast} + b\sqrt{P} - c,
\end{equation}
\begin{equation}
\frac{dD_{\rm st}^{\ast}}{dt} = I - \frac{D_{\rm st}^{\ast}}{\tau},
\end{equation}
where $D_{\rm st}^{\ast}$ is a correction to $D_{\rm st}$, $I$ represents the energy injection to the ring current, $\tau$ is the decay time of the ring current, $P$ is the solar wind dynamic pressure, and $b$ and $c$ are constants. The injection $I$ is driven by the solar wind dawn-dusk electric field, $vB_s$, where $v$ is the solar wind speed and $B_s$ is the southward component of the interplanetary magnetic field. Therefore, the solar wind speed and magnetic field are two key elements in the generation of geomagnetic storms. A major difference between the two approaches is that the decay time $\tau$ is constant in the formula of Burton et al.\cite{burton75} but varies with $vB_s$ in that of O'Brien \& McPherron\cite{om00}. This results in different $D_{\rm st}$ profiles as we see from Figure~5. Although both of the models have been successful in reproducing observed $D_{\rm st}$ values, they were largely drawn from geomagnetic storms with $D_{\rm st} > -200$ nT. Application of the formulas to extreme storms like the present one should be taken with caution, and one should compare different approaches to get a reliable conclusion.

\noindent\textbf{Acknowledgements} \\
The research was supported by the Recruitment Program of Global Experts of China under grant Y3B0Z2A99S, by the SPORT project under grant XDA04060801, by the Specialized Research Fund for State Key Laboratories of China, by the CAS/SAFEA International Partnership Program for Creative Research Teams, and by the STEREO project under grant NAS5-03131. C.M. has received funding from the European Union Seventh Framework Programme (FP7/2007-2013) under grant agreement No. 263252 [COMESEP], and was supported in part by a Marie Curie International Outgoing Fellowship within the 7th European Community Framework Programme. We acknowledge the use of data from STEREO and SOHO, and thank C. T. Russell and D. Odstrcil for helpful discussion.

\noindent\textbf{Author contributions} \\
Y. Liu designed the research and led the data analysis. All authors participated in the data analysis, interpretation, discussion and writing of the paper.  

\noindent\textbf{Additional information} \\
\textbf{Supplementary Information} accompanies this paper at http://www.nature.com/naturecommunications. \\\textbf{Competing financial interests:} The authors declare no competing financial interests.

\clearpage

\begin{figure}
\centerline{\includegraphics[width=30pc]{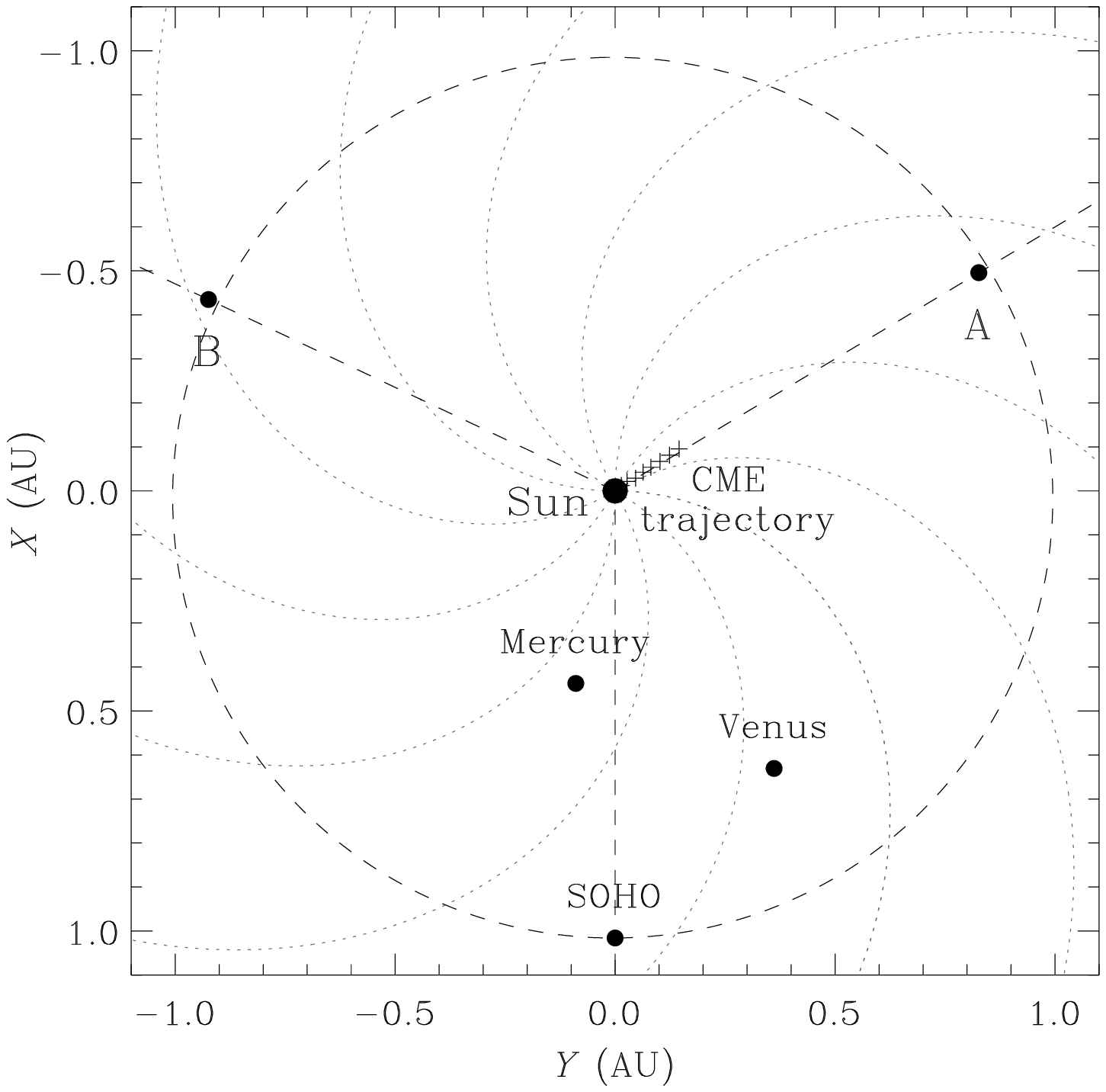}} 
{\textbf{Figure 1 $|$ Positions of the spacecraft and planets in the ecliptic plane on 2012 July 23.} The dashed circle indicates the orbit of the Earth, and the dotted lines show the spiral interplanetary magnetic fields. The trajectory of the merged CMEs as measured using SOHO and STEREO B coronagraph observations is marked by crosses.}
\end{figure}

\clearpage

\begin{figure}
\centerline{\includegraphics[width=38pc]{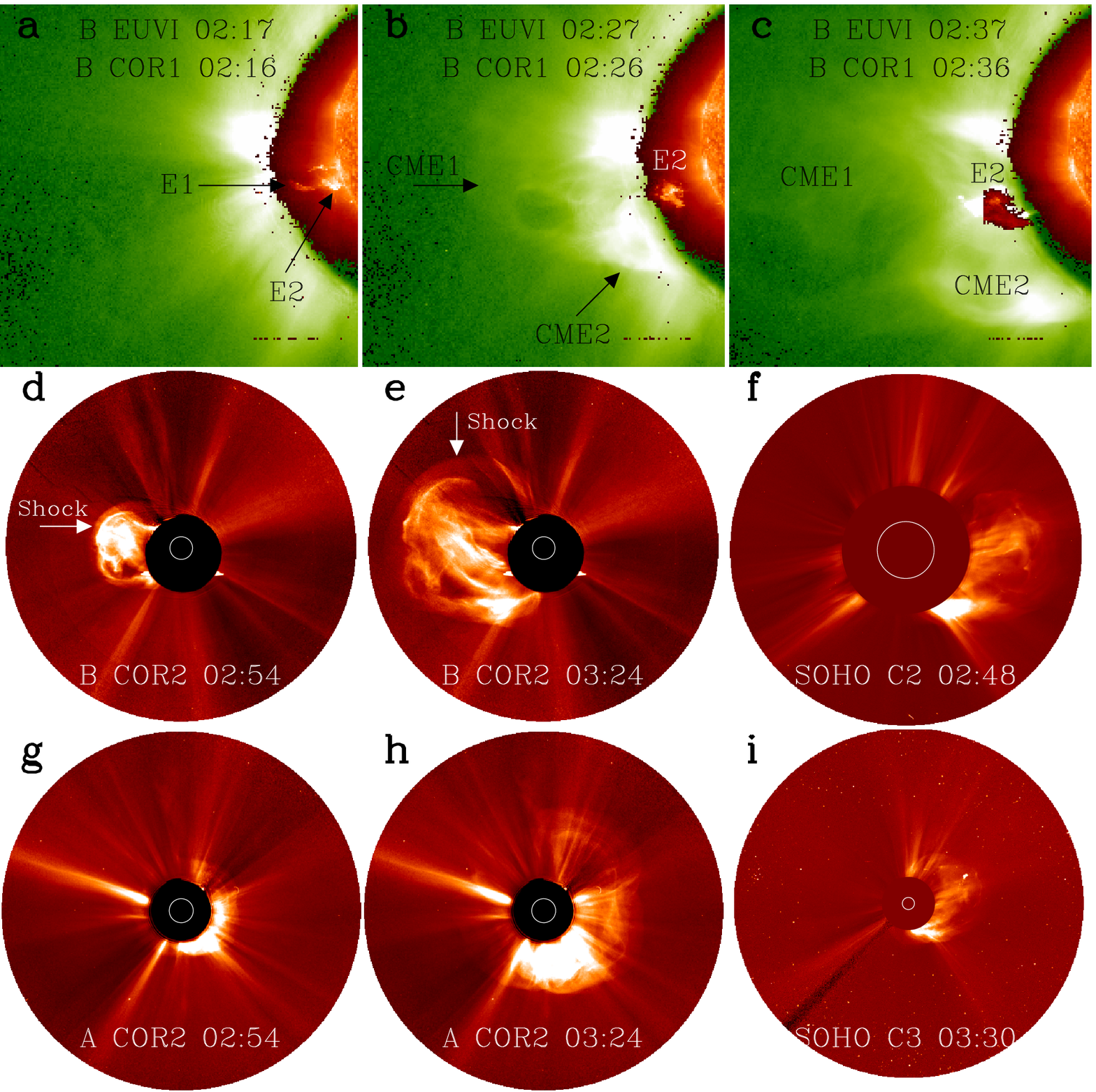}} 
{\textbf{Figure 2 $|$ Solar source and evolution of the 2012 July 23 event viewed from three vantage points.} (\textbf{a}-\textbf{c}) Composite images of EUVI at 304 \AA\ and COR1 aboard STEREO B, showing two consecutive prominence eruptions (E1 and E2) and corresponding nascent CMEs (CME1 and CME2). (\textbf{d}-\textbf{i}) Images from COR2 of STEREO B, COR2 of STEREO A, and LASCO of SOHO, respectively. Note a transition layer ahead of the front of the merged CMEs in COR2 of STEREO B, reminiscent of a shock signature. The white circle marks the location and size of the solar disk. Also see Supplementary Movies 1-3.}
\end{figure}

\clearpage

\begin{figure}
\centerline{\includegraphics[width=36pc]{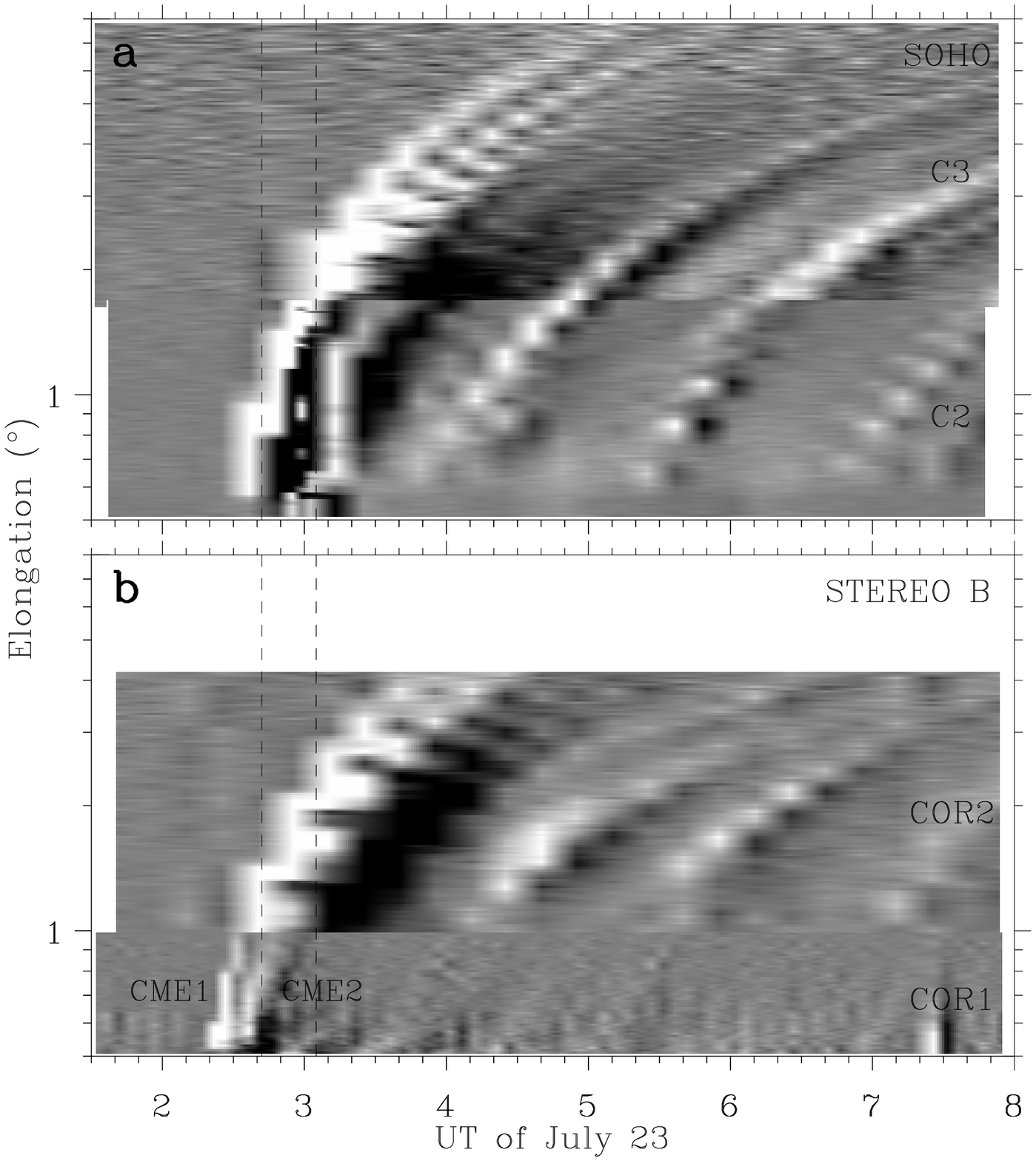}} 
{\textbf{Figure 3 $|$ Time-elongation maps constructed from running difference images along the ecliptic.} (\textbf{a}) Time-elongation maps from LASCO of SOHO. (\textbf{b}) Time-elongation maps from COR1 and COR2 of STEREO B. Two adjacent tracks (labeled as CME1 and CME2) can be seen in the map from COR1 of STEREO B, which is consistent with the scenario of two CMEs launched in quick succession. The two vertical dashed lines bracket the time period, during which the two tracks seem to merge and the change in the CME propagation direction occurred (see Fig.~4).}
\end{figure}

\clearpage

\begin{figure}
\centerline{\includegraphics[width=26pc]{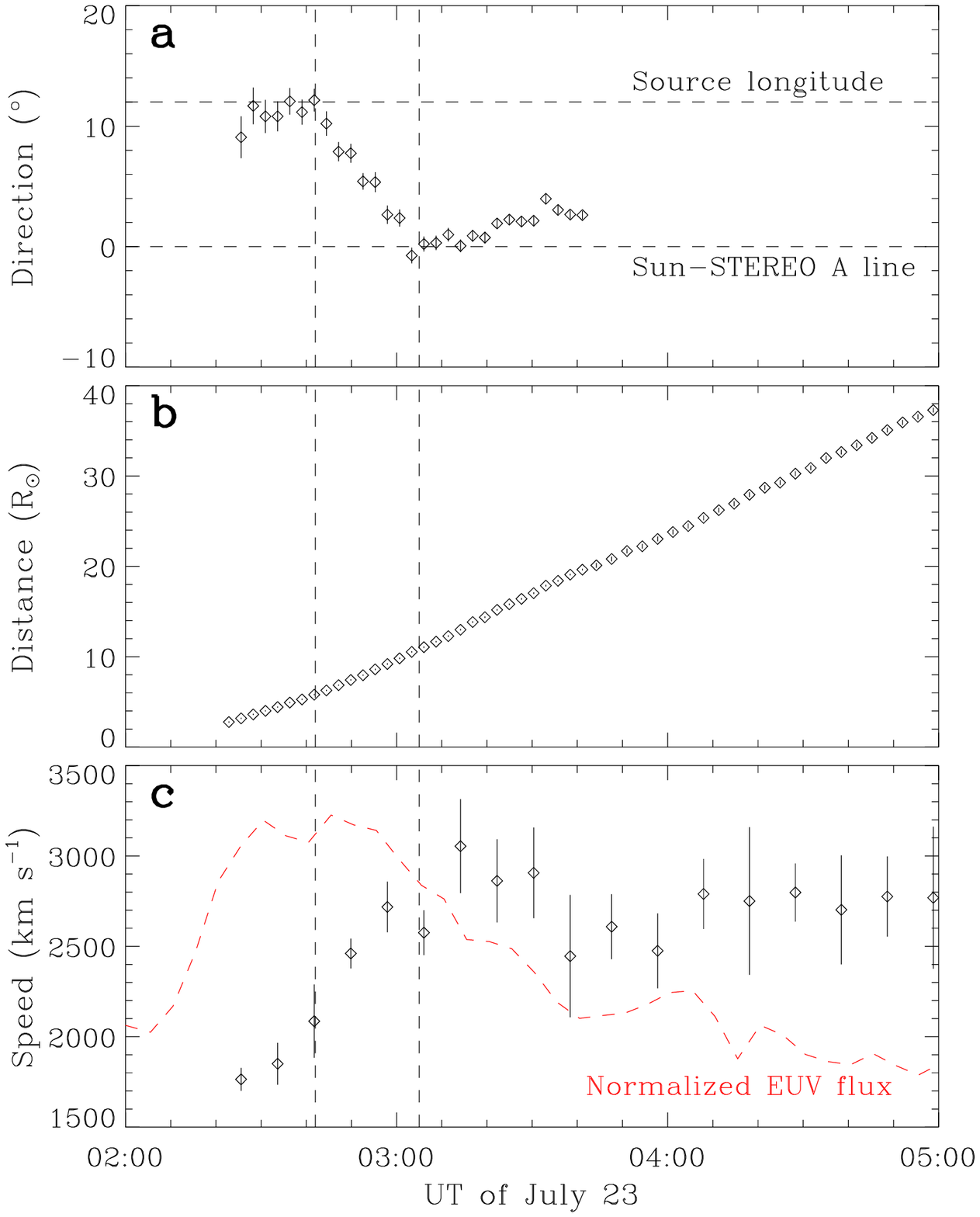}} 
{\textbf{Figure 4 $|$ Kinematics of the CME leading edge derived from a triangulation analysis.} (\textbf{a}) Propagation direction. (\textbf{b}) Radial distance. (\textbf{c}) Speed. The Sun-STEREO A line and the longitude of the solar source location are indicated by the horizontal dashed lines in panel (\textbf{a}). The two vertical dashed lines bracket the time interval during which the propagation angle shifted toward STEREO A. After 03:41 UT CME elongation measurements are available only from SOHO, so the distances after 03:41 UT are calculated from SOHO observations assuming a propagation direction of 2$^{\circ}$ west of STEREO A. The speeds are calculated from adjacent distances using a numerical differentiation technique. The error bars show uncertainties in the propagation angle, distance and speed. Overlaid on the speeds is the normalised full-disk EUV flux at 195 \AA\ observed by STEREO A (red curve; in arbitrary units).}
\end{figure}

\clearpage

\begin{figure}
\centerline{\includegraphics[width=27pc]{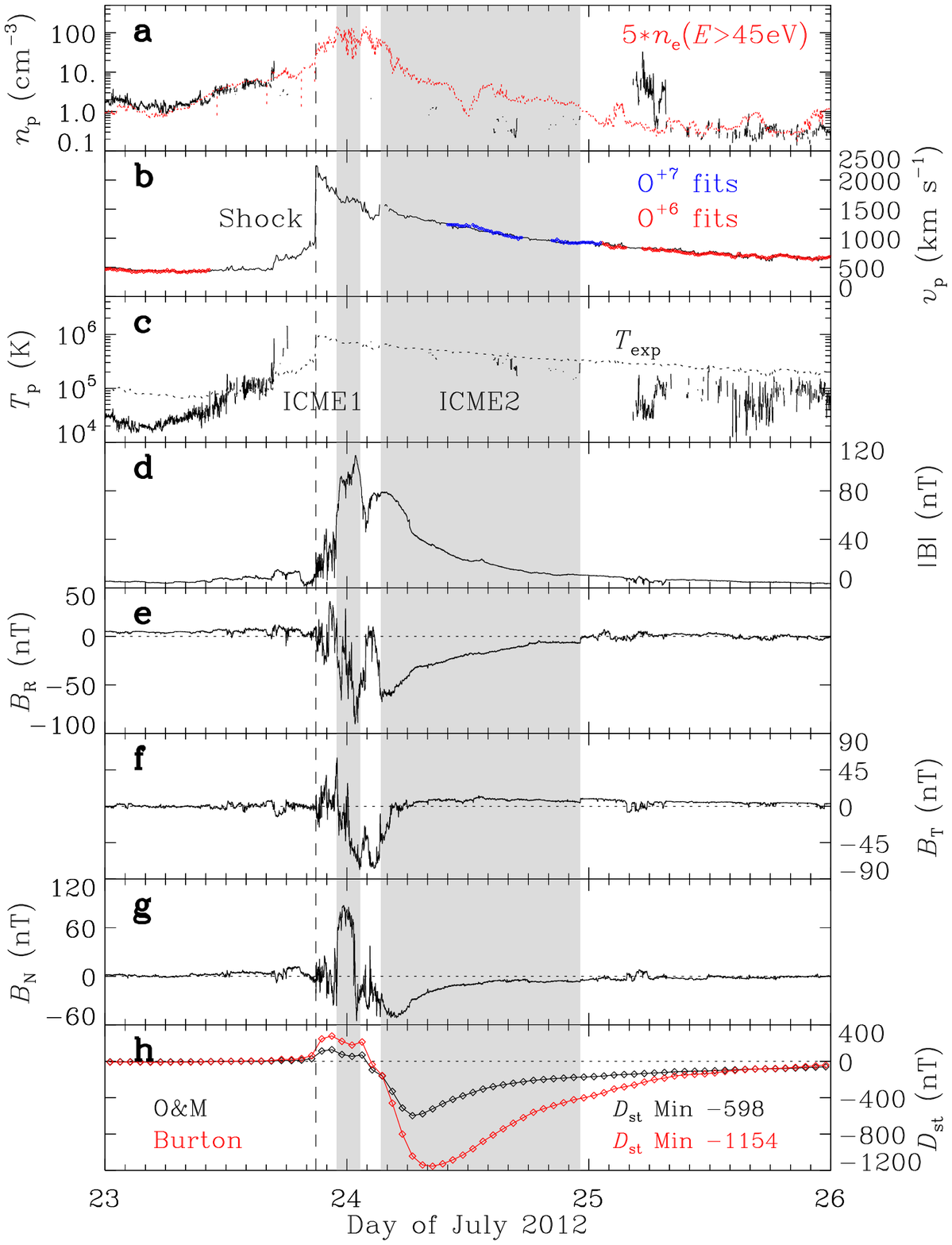}}
{\textbf{Figure 5 $|$ Solar wind parameters observed at STEREO A.} From top to bottom, the panels show the proton density (\textbf{a}), bulk speed (\textbf{b}), proton temperature (\textbf{c}), magnetic field strength (\textbf{d}) and components (\textbf{e}-\textbf{g}), and $D_{\rm st}$ index (\textbf{h}), respectively. The shaded regions indicate the ICME intervals, and the vertical dashed line marks the associated shock. The red curve in panel (\textbf{a}) represents the number density (multiplied by a factor of 5) of electrons with energies above 45 eV. Speeds derived from O$^{+6}$ and O$^{+7}$ fits are also shown in panel (\textbf{b}) and are consistent with the bulk speeds calculated from proton distributions. The dotted curve in panel (\textbf{c}) denotes the expected proton temperature from the observed speed. The $D_{\rm st}$ values in panel (\textbf{h}) are estimated using the formulas of Burton et al.\cite{burton75} (red) and O'Brien \& McPherron\cite{om00} (black), respectively.}
\end{figure}

\clearpage

\begin{figure}
\centerline{\includegraphics[width=30pc]{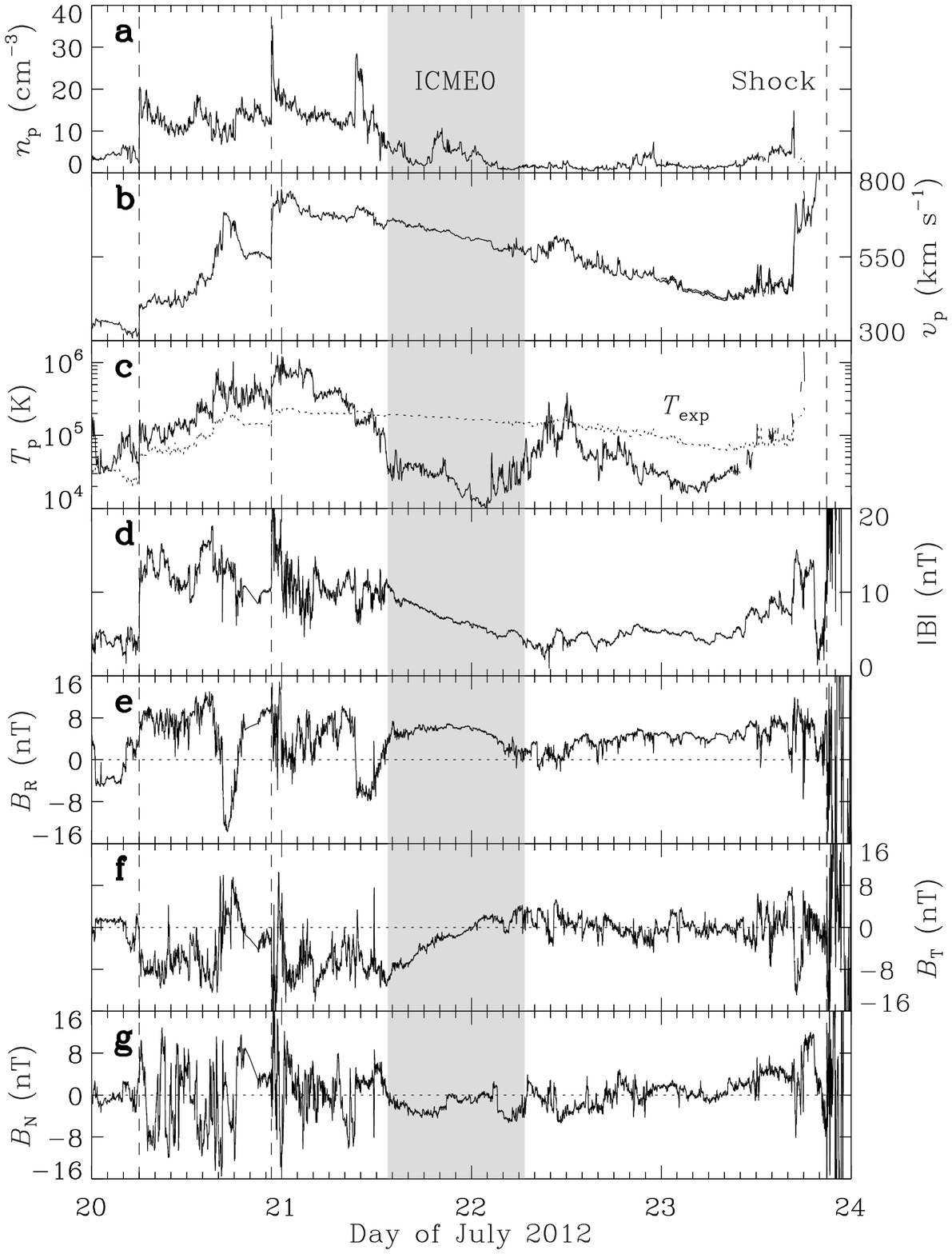}} 
{\textbf{Figure 6 $|$ Solar wind parameters observed at STEREO A before the 2012 July 23-24 event.} From top to bottom, the panels show the proton density (\textbf{a}), bulk speed (\textbf{b}), proton temperature (\textbf{c}), magnetic field strength (\textbf{d}) and components (\textbf{e}-\textbf{g}), respectively. The dotted curve in panel (\textbf{c}) denotes the expected proton temperature from the observed speed. An ICME (ICME0; shaded region) with two preceding shocks (vertical dashed lines) can be identified. There could be a second ICME region (or part of ICME0) on July 22-23 as indicated by the depressed proton temperature, but other ICME signatures are absent.}
\end{figure}

\clearpage

\begin{figure}
\centerline{\includegraphics[width=29pc]{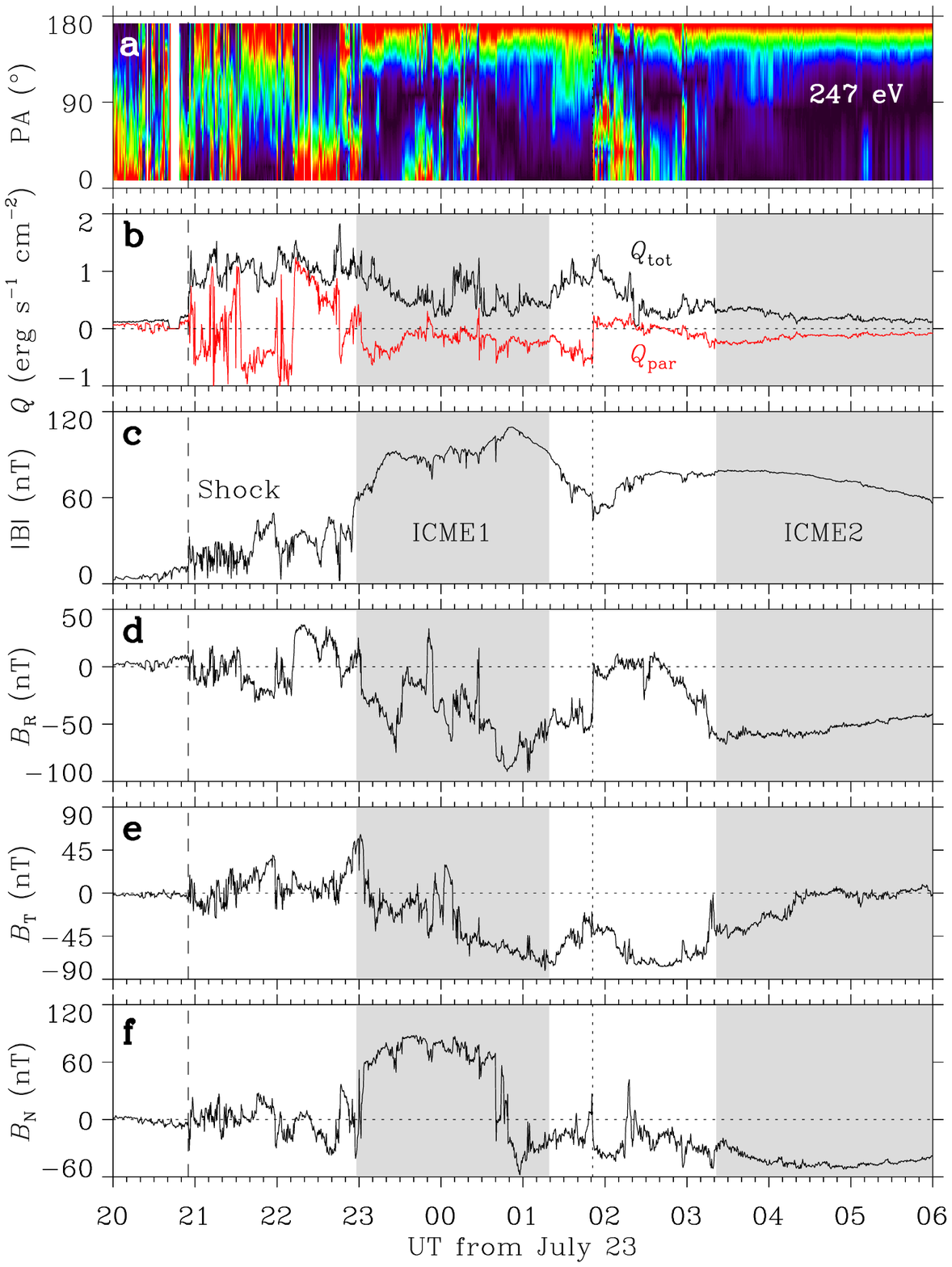}} 
{\textbf{Figure 7 $|$ Expanded view of the solar wind parameters across the shock and ICMEs.} (\textbf{a}) A normalised pitch-angle (PA) distribution of 247 eV electrons (with values descending from red to black). (\textbf{b}) Electron heat flux (black) and its component along the magnetic field (red). (\textbf{c}-\textbf{f}) Magnetic field strength and components, respectively. Both the PA distribution and heat flux are calculated in the spacecraft frame. The shaded regions denote the ICME intervals, and the vertical dashed line indicates the associated shock. The vertical dotted line marks a magnetic field polarity change in the interface of the CME-CME interaction.}
\end{figure}

\end{document}